\documentclass[aps,pre,twocolumn,showpacs,floatfix]{revtex4} 
\usepackage{amsmath,amssymb,isolatin1,graphicx,times}

\newcommand{\be}{\begin{equation}}
\newcommand{\ee}{\end{equation}}
\newcommand{\bea}{\begin{eqnarray}}
\newcommand{\eea}{\end{eqnarray}}
\newcommand{\bw}{\begin{widetext}}
\newcommand{\ew}{\end{widetext}}

\newcommand{\kommentar}[1]{}

\begin{document}
 
\title{Quantum transport on small-world networks: A
continuous-time quantum walk approach}
\author{Oliver M{\"u}lken}
\email{muelken@physik.uni-freiburg.de}
\author{Volker Pernice}
\author{Alexander Blumen}
\affiliation{
Theoretische Polymerphysik, Universit\"at Freiburg,
Hermann-Herder-Stra{\ss}e 3, 79104 Freiburg i.Br., Germany}

\date{\today} 
\begin{abstract}
We consider the quantum mechanical transport of (coherent) excitons on
small-world networks (SWN). The SWN are build from a one-dimensional ring
of $N$ nodes by randomly introducing $B$ additional bonds between them.
The exciton dynamics is modeled by continuous-time quantum walks and we
evaluate numerically the ensemble averaged transition probability to reach
any node of the network from the initially excited one. For sufficiently
large $B$ we find that the quantum mechanical transport through the SWN
is, first, very fast, given that the limiting value of the transition
probability is reached very quickly; second, that the transport does not
lead to equipartition, given that on average the exciton is most likely to
be found at the initial node.
\end{abstract}
\maketitle

\section{Introduction}

Many systems encountered in nature cannot be described by simple lattice
models. In general such systems are characterized by graphs whose bonds
connect sites with a wide distribution of mutual distances. Examples can
be found in various fields, ranging from physics or biology to social
studies or computer science; see
\cite{watts1998,albert2002,dorogovtsev2002} and references therein.  More
specifically, some of these systems can be described by small-world
networks (SWN), which have large clustering coefficients but short
characteristic path lengths \cite{albert2002}. The statistical properties
of SWN have been studied to a great extent and are now well understood. 

A large variety of dynamical processes on graphs are related to the
spectrum of the (discrete) Laplacian of the underlying topological network
\cite{alexander1982,bray1988,monasson1999}. For classical diffusion on
SWN, which has been modeled, for instance, by random walks
\cite{jespersen2000,jespersen2000b}, it was found that the
probability to be still or again at the initial site has a complex
dependence on the number $n$ of steps, i.e., at short times it decays as a
power-law of $n$, whereas at longer times it has a stretched exponential
dependence on $n$. The quantum dynamics on SWN has been studied mainly in
the framework of the localization-delocalization transition
\cite{zhu2000,giraud2005}, where one has also assumed an additional (on
site) disorder. Here, the transition depends on the complexity of the SWN.
A comparison between classical and quantum diffusion was given in
\cite{kim2003}, where a quantum diffusion time (defined as the time where
the participation ratio of the time-dependent wave function has dropped to
a certain value) was shown to be
faster than its classical counterpart. However, even here little
consideration has been given to the full set of eigenvectors of such
systems, which become important in the quantum mechanical extension of the
classical diffusion process. 

To be specific, a quantum mechanical analog of continuous-time random
walks (CTRW) can be defined by identifying the Laplacian (or connectivity
matrix) ${\bf A}$ of the network with the Hamiltonian ${\bf H}$. For
simple lattices this corresponds, in fact, to a nearest neighbor hopping
model \cite{farhi1998,childs2002,mb2005b,mbb2006a,bbm2006a}. The
transformation replaces the classical diffusion process by a quantal
propagation of the excitation through the network. Due to its formal
similarity to CTRW, the procedure was dubbed continuous-time quantum walk
(CTQW). In fact, it is known in other branches of physics under different
names, such as the tight-binding model in solid-state physics \cite{Ziman}
or the H{\"u}ckel/LCMO model in physical chemistry \cite{McQuarrie}.  CTQW
are also closely related to so-called quantum graphs (QG), see, for
instance, \cite{kottos1997, schanz2000, kottos2000, kottos2003}, whose
connectivity matrix is defined in a similar way. However, QG explicitly
consider the bond between two nodes in the sense that bonds may be
directed and are given a varying length. Thus, CTQW are, to some extent, a
simplified version of QG. Quite recently, Smilansky discussed the
connections between discrete Laplacians (equivalently, between the connectivity
matrices) on discrete QG and periodic orbits \cite{smilansky2007}.
There is certainly a large mathematical backbone on which to establish
further connections, see, for instance, \cite{kuchment2004}.

\section{Quantum walks on networks}

\begin{figure}[htb]
\centerline{\includegraphics[clip=,width=0.5\columnwidth]{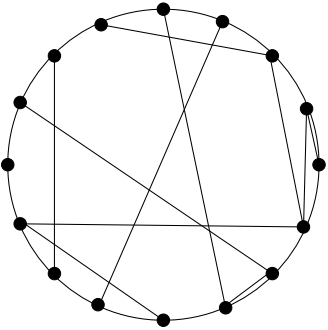}}
\caption{Sketch of a SWN of size $N=16$ containing $B=11$ additional bonds.}
\label{swn}
\end{figure}

Here, we consider transport processes (CTQW and CTRW) on networks, which
allows us to study the two extreme cases of transport processes on such
structures, namely, purely quantum mechanical (CTQW) and purely classical
processes (CTRW). Networks are a collection of $N$ connected nodes. The
periodicity of regular networks can be destroyed by randomly including $B$
additional bonds into the network.  In such a way one creates
``shortcuts'' and a walker can find shorter paths between pairs of sites
than on the regular network.  In the following we create the SWN by
randomly adding bonds to a regular one dimensional ring, see
Fig.~\ref{swn}. However, we forbid self-connections, i.e., bonds
connecting one node with itself.

We denote by $|j\rangle$ a state associated with a localized excitation at
node $j$ and take the set $\{|j\rangle\}$ to be orthonormal. For CTRW on
undirected and unweighted networks the transfer matrix is given by the
(discrete) Laplacian ${\bf A}$ of the network, by which we assume equal
transition rates $\gamma\equiv1$ between all nodes. The matrix ${\bf A}$
has as non-diagonal elements $A_{k,j}$ the values $-1$ if nodes $k$ and
$j$ of the network are connected by a bond and $0$ otherwise. The diagonal
elements $A_{j,j}$ of ${\bf A}$ equal the number of bonds $f_j$ which exit
from node $j$. Quantum mechanically, the states $|j\rangle$ span the whole
accessible Hilbert space; the time evolution of an excitation initially
placed at node $|j\rangle$ is determined by the systems' Hamiltonian ${\bf
H} = {\bf A}$ and reads $\exp(-i{\bf H}t)|j\rangle$, where we set
$\hbar\equiv1$. The classical and quantum mechanical transition
probabilities to go from the state $|j\rangle$ at time $0$ to the state
$|k\rangle$ in time $t$ are given by $ p_{k,j}(t) \equiv \langle k |
\exp(-{\bf A} t) | j \rangle $ and by $ \pi_{k,j}(t) \equiv
|\alpha_{k,j}(t)|^2 \equiv |\langle k | \exp(- i {\bf H} t) | j \rangle|^2
$, respectively. By fixing the coupling strength between two nodes
$|H_{j,j\pm1}|=1$, the time unit $[\hbar/H_{j,j\pm1}]$ for the transfer
between two nodes is set to unity.

From the eigenvalues $E_n$ of the Hamiltonian ${\bf H}$ (or Laplacian
${\bf A}$) follows the density of states (DOS or spectral density) of the
given system of size $N$,
\be
\rho(E) =
\frac{1}{N}\sum_{n=1}^N \delta(E-E_n).
\label{dos}
\ee
The DOS contains the essential information about the system and shows
distinct features which depend on the network's topology. These features
also carry over to dynamical properties, which in some cases depend only
on the $E_n$. For example, the {\sl average} classical probability to be
still or again at the initially excited node, 
\be
\overline{p}(t) =
\frac{1}{N} \sum_{n=1}^{N} \ e^{-E_n t},
\label{pclavg}
\ee
depends solely on the $E_n$ of ${\bf A}$, but {\sl not} on the eigenstates
$|\Phi_n\rangle$ \cite{bray1988}. In the quantum case, we find a lower
bound to  $\overline{\pi}(t) \equiv \frac{1}{N} \sum_{j=1}^{N}
\pi_{j,j}(t)$, which also depends only on the $E_n$
\cite{mb2006b,mbb2006a},
\be
\overline{\pi}(t) \geq
|\overline{\alpha}(t)|^2 = \left|\frac{1}{N} \sum_{n=1}^{N} \
e^{-iE_nt}\right|^2,
\label{pqmavg}
\ee
where $\overline{\alpha}(t) \equiv \frac{1}{N} \sum_{j=1}^{N}
\alpha_{j,j}(t)$. We hasten to note that the lower bound is exact for
regular networks \cite{mbb2006a,bbm2006a}. 
The quantity
$|\overline{\alpha}(t)|^2$ given in Eq.~(\ref{pqmavg}) has also been
derived in a different context as being the form factor of QG
\cite{kottos1997}.

\section{CTQW on SWN}

We 
will
analyze the general behavior of CTQW on SWN by averaging
over distinct realizations $R$
\be
\langle \cdots \rangle_R \equiv \frac{1}{R} \sum_{r=1}^R [\cdots]_r,
\ee
where the index $r$ specifies the $r$th realization of the quantity in
question. In so doing we obtain statistical results which allow for a
comparison with the classical ones. In particular, we will consider the
realization-averaged transition probabilities
$\langle\pi_{kj}(t)\rangle_R$, the averaged probabilities
$\langle\overline{\pi}(t)\rangle_R$, their lower bound
$\langle\overline{\alpha}(t)\rangle_R$, and their classical analog
$\langle\overline{p}(t)\rangle_R$. Furthermore, we also calculate the long
time average (LTA) of each quantity:  
\be
\left\langle \lim_{T\to\infty} \frac{1}{T} \int_0^T dt \
\cdots \right\rangle_R.
\ee

For the numerical evaluation we make
use of the standard software package MATLAB. Specifically, we focus on SWN
of size $N=100$ with $B=1$, $2$, $5$, and $100$ additional bonds; the
ensemble average is, in general, performed over $R=500$ realizations,
which guarantees a sufficiently large number of samples under manageable
computing times.

\subsection{Random matrix theory}

Before going into the details of our analysis, we like to point to the
differences and similarities of SWN with other approaches to study quantum
transport processes.  Classical transport over SWN differs from that over
other systems, such as regular lattices or fractal networks, it that the
transport gets to be faster: While the probability to return to the
origin decays as $t^{-1/2}$ for
regular networks, it decays as a stretched exponential for SWN
\cite{jespersen2000,jespersen2000b}, vide infra Fig.~\ref{prob_kk}(a).
While the classical dynamics over SWN is by now well-understood, little is
known about the quantum dynamics on such networks. 

In general, several dynamical properties of networks depend only on the
DOS of the system's Hamiltonian \cite{Cvetkovic}. We choose the additional
bonds of our SWN randomly, thus the corresponding Hamiltonian will have
entries at random positions in the matrix. This has to be distinguished
(to some extent) from random matrix theory (RMT) \cite{Mehta}. However,
there are also similarities between RMT and SWN. The DOS of SWN have been
compared to RMT in \cite{bandyopadhyay2007}, where it was found that the
level spacing $\Delta E\equiv(E_{n+1}-E_{n})$ of the DOS of SWN can be
fitted by the so-called Brody distribution,
which interpolates
between Poissonian 
and Wigner-Dyson
level spacings
statistics, see \cite{bandyopadhyay2007} for details. The SWN considered in
Ref.~\cite{bandyopadhyay2007} is a Watts-Strogatz network, obtained by
randomly permuting the bonds of a regular one-dimensional network. The
eigenvalue statistics of random networks have been studied in
Ref.~\cite{mirlin2000} and in the works referenced therein; 
the quantum dynamics on regular disordered networks has been considered in
\cite{klesse1999}.

\bw
~\\
\begin{figure}[htb]
\centerline{\includegraphics[clip=,width=\columnwidth]{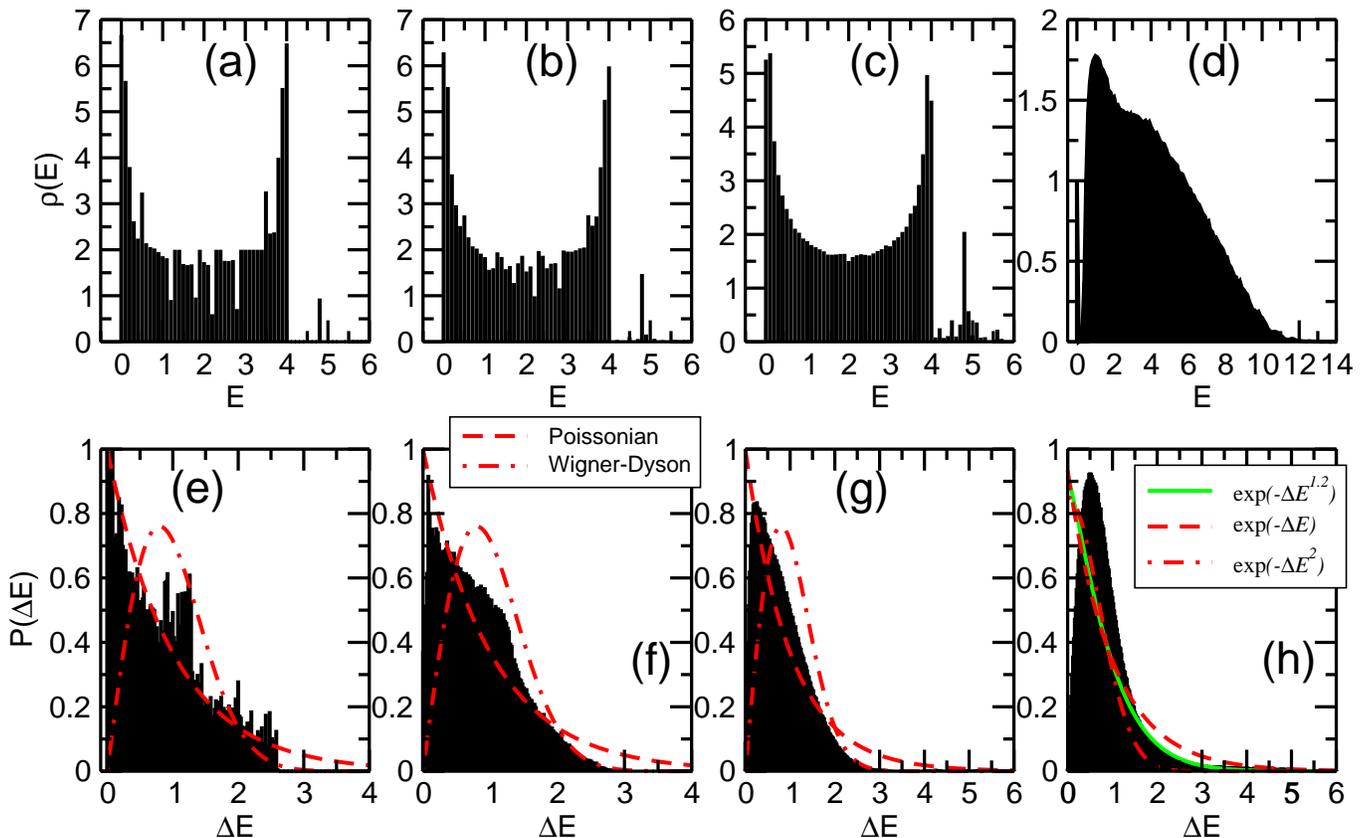}}
\caption{(Color online) DOS $\rho(E)$ (a)-(d) and level spacing
distribution $P(\Delta E)$ (e)-(f) of SWN with $N=100$ nodes and $B=1$
[(a),(e)], $2$ [(b),(f)], $5$ [(c),(g)], and $100$ [(d),(h)] additional
bonds. The lower panels (e)-(g) show also the Poissonian (dashed line) and
Wigner-Dyson (dashed-dotted line) statistics, panel (h) shows fits of the
tails of $P(\Delta E)$ with different exponentials.}
\label{swn_dos}
\end{figure}
\ew

Now, the DOS of a SWN differs from that of networks whose sites have been
totally randomly connected; the DOS of the latter networks follow Wigner's
semicircle law. Figure \ref{swn_dos} shows for SWN with $N=100$ nodes and
$B=1$, $2$, $5$, and $100$ additional bonds histograms of the (average)
DOS $\rho(E)$ and of the level spacing distribution $P(\Delta E)$, where
$\Delta E$ is normalized in such a way that the average
$\overline{\Delta E}=1$. While for small $B$ the DOS barely changes, the
level spacing distribution shows more drastic changes, see
Fig.~\ref{swn_dos}~(a)-(c). 
The appearance of large isolated eigenvalues results in a non-vanishing
$P(\Delta E)$ for large $\Delta E$. In
Figs.~\ref{swn_dos}~(e)-(h) [plots of $P(\Delta E)$] we also show the
Poissonian [$\exp(-\Delta E)$, dashed line]
and Wigner-Dyson \{$2\Gamma(3/2)^2\Delta E \exp[-\Gamma(3/2)^2 \Delta
E^2]$, dashed-dotted line\} statistics. While
$P(\Delta E)$ roughly follows the Poissonian statistics for $B=1$
[Fig.~\ref{swn_dos}~(e)], this is not the case when increasing $B$.
Especially the tail of the distribution $P(\Delta E)$ is better fitted by
the Wigner-Dyson statistics [Figs.~\ref{swn_dos}~(f) and (g)].  However,
when increasing $B$ to the order of $N$ [Fig.~\ref{swn_dos}~(h)], the tail
of $P(\Delta E)$ neither decays as $\exp(-\Delta E)$ (dashed line) nor as
$\exp(-\Delta E^2)$ (dashed-dotted line), but rather as $\exp(-\Delta
E^\mu)$, with $\mu\approx1.2$ (solid line).  Thus, the complexity of the
DOS of SWN (compared, e.g., to the semicircle law) leads to dynamical
properties of the SWN not all of which can be captured by RMT.

\subsection{Transition probabilities}\label{sec_transprob}

\begin{figure}[htb]
\centerline{\includegraphics[clip=,width=\columnwidth]{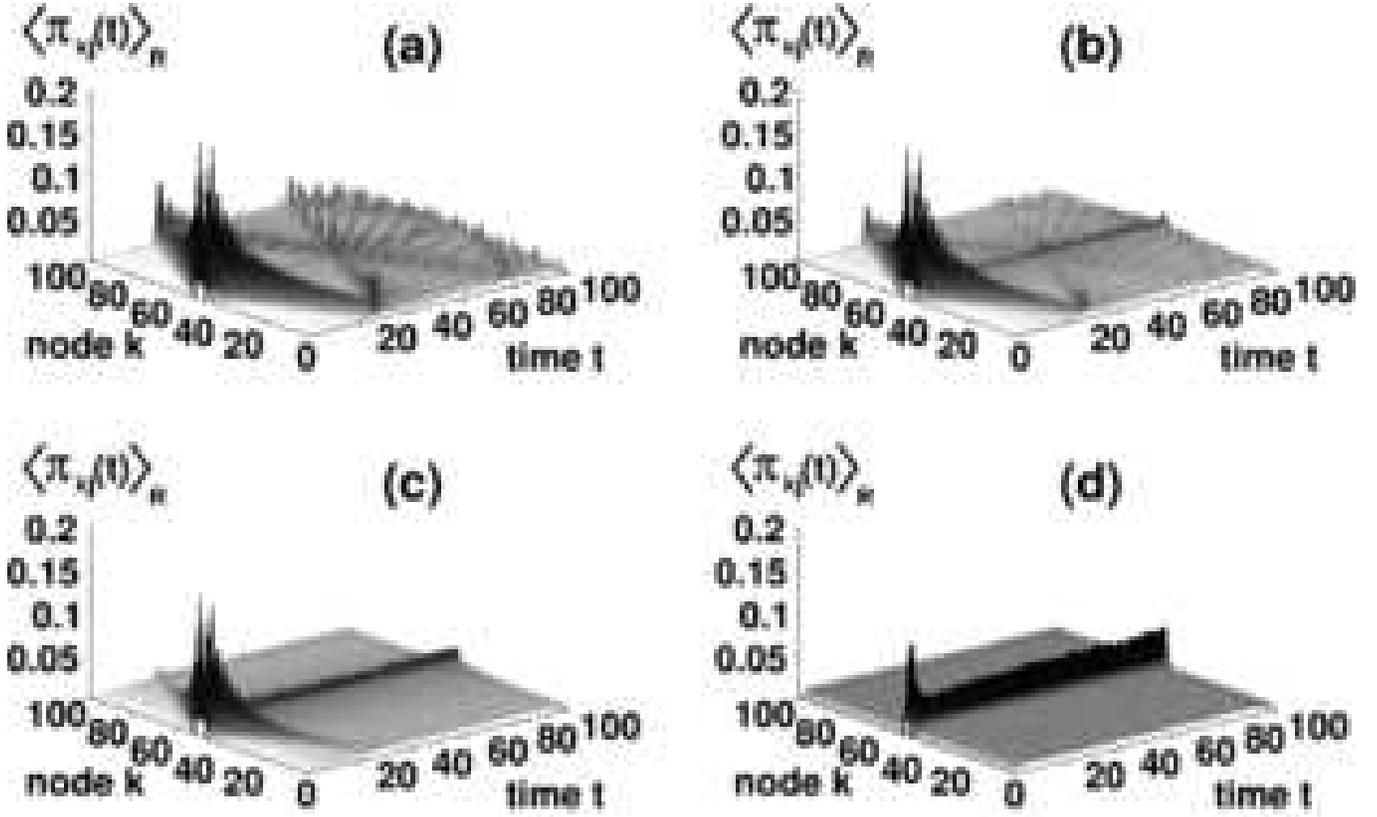}}
\caption{Time dependence of the averaged transition probabilities
$\langle\pi_{kj}(t)\rangle_R$ for SWN of size $N=100$ with (a) $B=1$, (b)
$B=2$, (c) $B=5$, and (d) $B=100$. The initial node is $j=50$ and the
number of realizations is $R=500$.}
\label{pikj}
\end{figure}

The ensemble average of the transition probabilities
$\langle\pi_{kj}(t)\rangle_R$ allows a first glimpse on the behavior of
CTQW on SWN. Figure \ref{pikj} shows $\langle\pi_{kj}(t)\rangle_R$ for
several SWN with $N=100$ nodes and different $B$. Note that due to the
ensemble average we can choose the initial node $j$ freely, and we thus
take $j=50$. In the absence of any additional bond, the excitations travel
along the ring and interfere in a very regular manner, producing discrete
quantum carpets \cite{mb2005b}. Typical for these carpets is that they
show, depending on $N$, full or partial revivals at specific times
\cite{mb2005b}.

\bw
~
\ew

For SWN the situation is quite different. Already a few additional bonds
obliterate the quantum carpets; the patterns fade away. By adding more
bonds, only the initial node retains a significant value for
$\langle\pi_{jj}(t)\rangle_R$ at all times $t$. Furthermore, already for
SWN with as little as $B=5$ the pattern of $\langle\pi_{jj}(t)\rangle_R$
becomes quite regular after a short time, see Fig.~\ref{pikj}(c). 
This
almost regular shape is reached very quickly when $B$ gets to be
comparable to $N$ [Fig.~\ref{pikj}(d)]. We note, however, that particular
realizations may still show (depending on their actual additional bonds)
strong interference patterns. These features are washed out by the
ensemble average, so that only the dependence on the initial node stands
out. 
We will return to the discussion of the transition probabilities
$\langle\pi_{kj}(t)\rangle_R$ in Sec.\ \ref{sec_partratio}.

For the ring the LTA can be calculated analytically. Depending on whether
$N$ is even or odd, the LTA are slightly different \cite{mb2005b}.  For
even $N$ (superscript $^e$) there are two maxima at $k=j$ and at $k=j+N/2$, both having the
value $\chi^e_{k,j} \equiv \lim_{T\to\infty} \frac{1}{T} \int_0^T dt \
\pi_{k,j}(t)= (2N-2)/N^2$; this is due to the fact that the number
of nodes from $j$ to $j+N/2$ is the same in both directions, which leads
to constructive interference. On the other hand, for odd $N$ (superscript
$^o$) there is only
one maximum at $k=j$, $\chi^o_{k,j} = (2N-1)/N^2$.

\begin{figure}[htb]
\centerline{\includegraphics[clip=,width=\columnwidth]{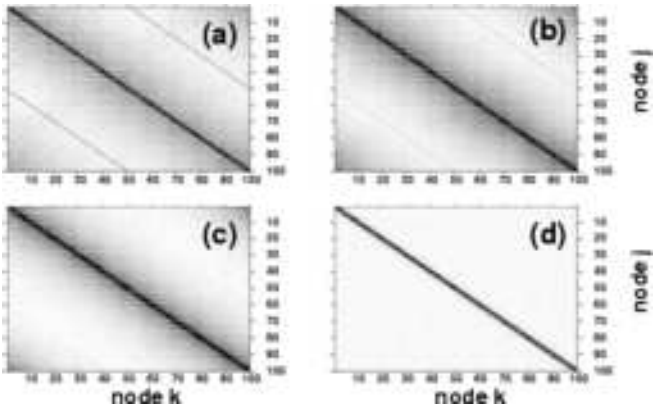}}
\caption{Long time average $\langle\chi_{k,j}\rangle_R$ for SWN of size
$N=100$ with (a) $B=1$, (b) $B=2$, (c) $B=5$, and (d) $B=100$. The
number of realizations is $R=500$. Dark regions denote large values of
$\langle\chi_{k,j}\rangle_R$ and bright regions low values of
$\langle\chi_{k,j}\rangle_R$.}
\label{chi}
\end{figure}

Figure \ref{chi} shows $\langle\chi_{k,j}\rangle_R$ for SWN of size
$N=100$ with $B=1$, $2$, $5$, and $100$. For $B=1$ and fixed $j$, the two
peaks of the regular network turn into a main peak and into a much weaker
side peak at $k=j+N/2$. This structure is still (barely) visible for
$B=2$. Already for $B=5$ the side peak has practically vanished; see
Fig.~\ref{chi}(c). While for $B=1,2$ and $5$ also structure around the
main peak is visible, for $B=100$, the $\langle\chi_{k,j}\rangle_R$ are
sharply peaked at $k=j$.  We stress that this should not be confused with
the Anderson localization, since there is a non-vanishing probability to
go from node $j$ to all other nodes $k\neq j$. The sharp peak of
$\langle\pi_{jj}(t)\rangle_R$ at the initial node $j$ is only the result
of ensemble averaging.

\subsection{Return probabilities}

Since CTQW on SWN always carry the information of their initial node $j$,
the averaged probabilities to return to $j$ are a good measure to quantify
the efficiency of the transport on such networks \cite{mb2006b}.

\begin{figure}[htb]
\centerline{\includegraphics[clip=,width=\columnwidth]{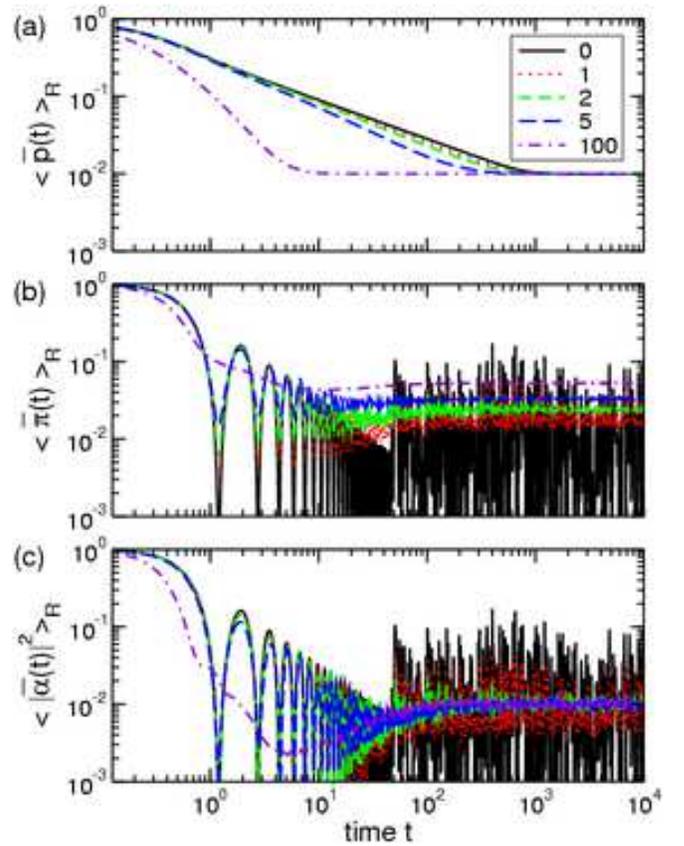}}
\caption{(Color online) Time dependence of the averaged probabilities (a)
$\langle\overline{p}(t)\rangle_R$, (b)
$\langle\overline{\pi}(t)\rangle_R$, and (c)
$\langle|\overline{\alpha}(t)|^2\rangle_R$ for SWN of size $N=100$ with
$B=1$, $2$, $5$, and $100$. The number of realizations is $R=500$.}
\label{prob_kk}
\end{figure}

Figure \ref{prob_kk} shows in double-logarithmic scales the ensemble
averages $\langle\overline{p}(t)\rangle_R$,
$\langle\overline{\pi}(t)\rangle_R$, and
$\langle\overline{\alpha}(t)\rangle_R$ for SWN with $N=100$ nodes and
$B=1$, $2$, $5$, and $100$. For classical transport
[Fig.~\ref{prob_kk}(a)] the initial decay of
$\langle\overline{p}(t)\rangle_R$ occurs faster for larger $B$.  The decay
at intermediate times follows a power-law ($t^{-1/2}$) for the ring (as is
clear from the linear behavior in the scales of the figure) and changes to
a stretched exponential-type when $B$ is large \cite{jespersen2000}. Thus,
a classical excitation will quickly explore the whole SWN, so that it will
occupy each site with equal probability of $1/N$ already after a
relatively short time, see the final plateau in Fig.~\ref{prob_kk}(a). 

\begin{figure}[htb]
\centerline{\includegraphics[clip=,width=\columnwidth]{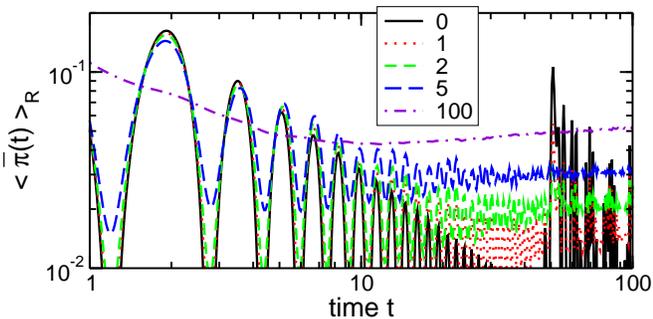}}
\caption{(Color online) Zoom into Fig.~\ref{prob_kk}(b) for short times
$t=1,\dots, 100$.}
\label{prob_kk_zoom}
\end{figure}

Quantum mechanically, however, the situation is more complex. Let us start
with the ensemble average $\langle\overline{\pi}(t)\rangle_R$, shown in
Fig.~\ref{prob_kk}(b). For a ring of $N$ nodes and for times smaller than
roughly $N/2$ $\langle\overline{\pi}(t)\rangle_R$ displays a quasiperiodic
pattern (black curve), the maxima of which decay as $t^{-1}$. At longer
times interference sets in and leads to an irregular behavior at times
larger than $N/2$ \cite{mb2006b}. Now, for SWN, as long as $B$ is
considerably less than $N$, the periodic pattern still remains visible; in
Fig.~\ref{prob_kk}(b) one can follow how an increase in $B$ (red, green,
and blue curves) is smoothing out the curves, so that both the heights of
the first maxima and the depths of the minima decrease.  At longer times
the SWN patterns are flattened out and $\langle\overline{\pi}(t)\rangle_R$
tends towards a limiting value.  With increasing $B$ this asymptotic
domain is reached more quickly. To emphasize this point we display in
Fig.~\ref{prob_kk_zoom} in an enlarged scale the data of
Fig.~\ref{prob_kk}(b) in the time interval $[1,100]$. Clearly, for larger
$B$ the crossover from the quasiperiodic behavior at short times to a
smoothed out pattern at longer times is shifted to smaller $t$. 

In  Fig.~\ref{prob_kk}(c) we plot the lower bound of $\overline{\pi}(t)$,
namely $\langle|\overline{\alpha}(t)|^2\rangle_R$ averaged over the
realizations. We note that the overall behavior of Figs.~\ref{prob_kk}(b)
and \ref{prob_kk}(c) is quite similar. However, the limiting values at
long times differ. For the LTA of $\langle\overline{\pi}(t)\rangle_R$ we
have (see also Eq.~(17) of Ref.~\cite{mvb2005a})
\bea
\langle\overline{\chi}\rangle_R 
&\equiv& \Big\langle
\lim_{T\to\infty}\frac{1}{T} \int_0^T dt \ \overline{\pi}(t) \Big\rangle_R 
\nonumber \\
&=& \frac{1}{RN} \sum_{r,j,n,n'}
\delta(E_{n,r}-E_{n',r}) \big|\langle j | \Phi_{n,r} \rangle \langle j |
\Phi_{n',r} \rangle \big|^2,
\label{pikk_avg}
\eea
where $\delta(E_{n,r}-E_{n',r})=1$ for $E_{n,r}=E_{n',r}$ and
$\delta(E_{n,r}-E_{n',r})=0$ otherwise.  For
$\langle|\overline{\alpha}(t)|^2\rangle_R$ the long-time values for
different $B$ collapse to one value. In fact, the LTA of
$\langle|\overline{\alpha}(t)|^2\rangle_R$ obeys
\be
\Big\langle
\lim_{T\to\infty}\frac{1}{T} \int_0^T dt \ |\overline{\alpha}(t)|^2
\Big\rangle_R 
= \frac{1}{RN^2} \sum_{r,n,n'} \delta(E_{n,r}-E_{n',r}), 
\label{alphakk_avg}
\ee
as can be immediately inferred from Eq.~(\ref{pqmavg}). Thus this quantity
is only a function of the eigenvalues $E_{n,r}$ and does not depend on the
eigenstates $|\Phi_{n,r}\rangle$.  In order to quantify the differences
between Eqs.~(\ref{pikk_avg}) and (\ref{alphakk_avg}) for SWN, we will
assume that all the eigenvalues are nondegenerate (this assumption is, of
course, not valid for the ring, see below). In Eq.~(\ref{alphakk_avg}) the
triple sum adds then to $RN$, so that the rhs equals $1/N$. On the other
hand, Eq.~(\ref{pikk_avg}) leads to 
\be
\langle\overline{\chi}\rangle_R = \frac{1}{RN}  \sum_{r,j,n}
\big|\langle j |
\Phi_{n,r} \rangle \big|^4. 
\label{chi_avg}
\ee
This expression depends on the eigenstates; in fact the rhs of
Eq.~(\ref{chi_avg}) is the ensemble average of the averaged participation
ratio of the eigenstates $| \Phi_{n,r} \rangle$.  Equation (\ref{chi_avg})
is well known in the theory of quantum localization, see, e.g.,
Sec.~V.~A.\ in \cite{heller1987}.

\begin{figure}[htb]
\centerline{\includegraphics[clip=,width=0.9\columnwidth]{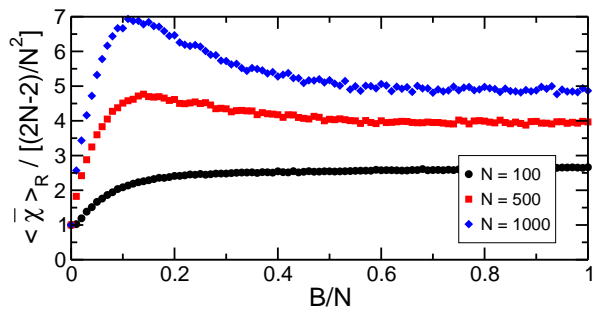}}
\caption{(Color online) The LTA of $\langle\overline{\pi}(t)\rangle_R$,
$\langle\overline{\chi}\rangle_R$, for SWN with $N=100$, $500$, and $1000$
nodes as a function of $B/N$.}
\label{pikk_lim}
\end{figure}

Now, Fig.~\ref{pikk_lim} shows the behavior of
$\langle\overline{\chi}\rangle_R$, according to Eq.~(\ref{pikk_avg}), for
a SWN with $N=100$, $500$, and $1000$ nodes as a function of $B/N$ (we
restrict ourselves to even $N$, the case of odd $N$ is similar).
Increasing $B$ results in an increase of
$\langle\overline{\chi}\rangle_R$, starting from the corresponding value
for the ring ($B=0$, only one realization, and $N$ even)
\be
\langle\overline{\chi}_{\rm ring}\rangle_R \equiv \overline{\chi} = \frac{1}{N} \sum_{j} \chi_{jj} =
\frac{2N-2}{N^2},
\label{pikk_avg_ring}
\ee 
where $\chi_{jj} = (2N-2)/N^2$. From Eq.~(\ref{alphakk_avg}) we obtain a
$1/N$ dependence for the LTA of
$\langle|\overline{\alpha}(t)|^2\rangle_R$, which by rescaling with
$\langle\overline{\chi}_{\rm ring}\rangle_R\sim 1/N$ would result in a
constant value for large $N$.  However, rescaling
$\langle\overline{\chi}\rangle_R$ with $\langle\overline{\chi}_{\rm
ring}\rangle_R$ shows an increase with $N$ of
$\langle\overline{\chi}\rangle_R/\langle\overline{\chi}_{\rm
ring}\rangle_R$ which is less than linear, thus,
$\langle\overline{\chi}\rangle_R$ depends on $N$ as $1/N^\nu$, with
$\nu\in[1,2]$. Additionally,  for larger $N$ (see $N=500$ and $1000$),
$\langle\overline{\chi}\rangle_R$ has a maximum value at $B/N \approx
0.14$, which is not present for smaller $N$ (see $N=100$), meaning that
for this ratio of $B/N$ the transport from the initial node to all others
is least probable, a fact which remains unclear. A detailed study of the
$N$ dependence will be given elsewhere.  When increasing $B$ to the order
of $N$, $\langle\overline{\chi}\rangle_R$ saturates to a plateau which
increases monotonically with $N$. Thus, an increase in the number of nodes
leads to a less probable transport from the initial node to all others.

We further note that with increasing $B$ the structures of
$\langle|\overline{\alpha}(t)|^2\rangle_R$ and
$\langle\overline{\pi}(t)\rangle_R$ differ even at short times, while for
the ring the relation $\overline{\pi}(t) = |\overline{\alpha}(t)|^2$ holds
exactly.

In Ref.~\cite{mb2006b} we showed that $\langle\overline{p}(t)\rangle_R$
and $\langle\overline{\pi}(t)\rangle_R$ (or
$\langle|\overline{\alpha}(t)|^2\rangle_R$) can be regarded as measures
for the efficiency of the excitonic transport. When increasing $B$, the
initial quantum transport through the SWN takes place - on average -
during a very short time scale (see Fig.~\ref{pikj}) compared to the ring,
where an excitation takes about $t=N/2$ to travel around the ring
\cite{mb2005b}. Additionally and in contrast to the classical case, where
the limiting value is always given by the equipartition value $1/N$, for
CTQW the limiting probability to be still or again at the initial node
increases with $B$. Thus, an exciton is (on average) more likely to be
found at the initial node, a feature which is not captured by the lower
bound $\langle|\overline{\alpha}(t)|^2\rangle_R$.  Therefore,
$\langle|\overline{\alpha}(t)|^2\rangle_R$ [as, for instance, shown in
Fig.~\ref{prob_kk}(c)] does not capture fine details of the transport,
which the full expression $\langle\overline{\pi}(t)\rangle_R$ does.

\subsection{Participation ratio of eigenstates}\label{sec_partratio}

\begin{figure}[htb]
\centerline{\includegraphics[clip=,width=\columnwidth]{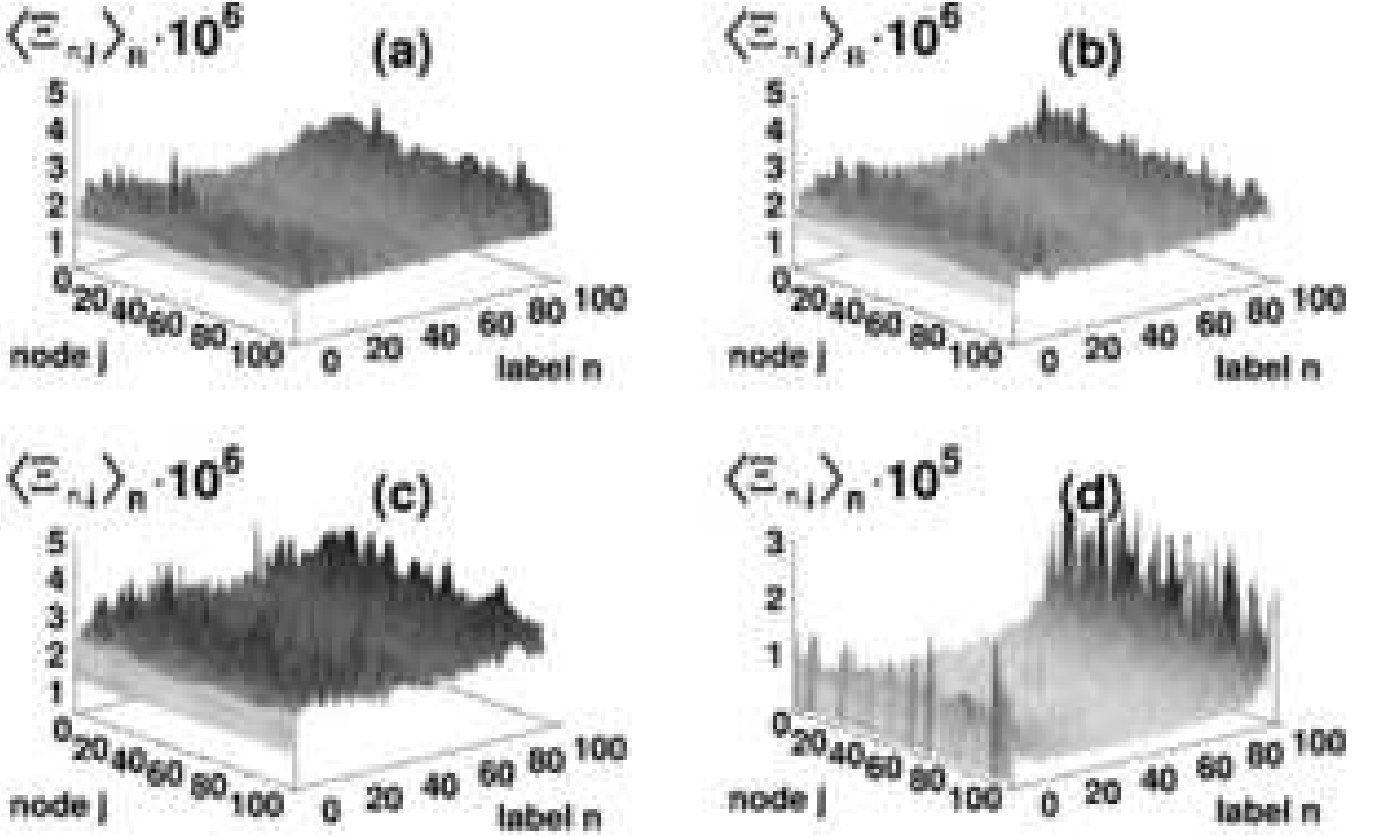}}
\caption{The function $\langle
\Xi_{n,j} \rangle_R$, Eq.~(\ref{partratio_eq}), for SWN of size $N=100$ with (a) $B=1$, (b)
$B=2$, (c) $B=5$, and (d) $B=100$. Note the different scaling
of the $z$-axis in (d). The number of realizations
is $R=500$.}
\label{partratio}
\end{figure}

For the ring the eigenstates are Bloch states,
\be
|\Phi_n \rangle = \frac{1}{\sqrt{N}} \sum_{j=1}^N e^{ iE_n j} |j\rangle,
\label{bloch}
\ee
from which $\big|\langle k | \Phi_{n} \rangle \big|^4 = 1/N^2$ follows for
all $|\Phi_{n} \rangle$. 
By naively inserting this result into
Eq.~(\ref{chi_avg}) one obtains $\langle\overline{\chi}\rangle_R = 1/N$,
which differs from the exact result, Eq.~(\ref{pikk_avg_ring}), by a
factor of 2. The reason for this difference is that for a ring most of the
eigenvalues are doubly degenerate. For SWN, on the other hand, most
eigenvalues are non-degenerate. The fact that, as is evident from
Fig.~\ref{pikk_lim}, $\langle\overline{\chi}\rangle_R$ for SWN increases
with increasing $B$ points towards a change of the $\big|\langle k |
\Phi_{n} \rangle \big|^4$ from the value $1/N^2$. In order to quantify the
difference to the ring case we plot in Fig.~\ref{partratio} the average
distribution of eigenstates, 
\be
\langle \Xi_{n,j} \rangle_R \equiv \frac{1}{RN} \sum_{r} \big| \langle j | \Phi_{n,r} \rangle
\big|^4 
\label{partratio_eq}
\ee
for SWN with $N=100$ with $B=1$, $2$, $5$, and $100$. From
Fig.~\ref{partratio} we remark that the $\langle \Xi_{n,j} \rangle_R$
increase with increasing $B$.  Additionally, the fluctuations between
different values of $\langle \Xi_{n,j} \rangle_R$ become larger, too. This
results in a substantial increase of $\langle\overline{\chi}\rangle_R$ for
larger $B$. We stress the particular role played by the eigenstate
$|\Phi_0\rangle = N^{-1/2}\sum_j |j\rangle$, which corresponds to the
eigenvalue $E_0=0$ and for which $\langle \Xi_{0,j} \rangle_R = 1/N^3$.
Most of the other states contribute more to
$\langle\overline{\chi}\rangle_R$. In particular for SWN with large $B$,
Fig.~\ref{partratio}(d), one finds large values for $\langle \Xi_{n,j}
\rangle_R$ close to the band edges of $E_n$ (i.e., for $n$ close to $0$
and close to $N$), in accordance with previous work; see, for instance
Ref.~\cite{farkas2001}.

The situation may be visualized as follows: For the ring all eigenstates
are Bloch states and hence are completely delocalized. Going over to SWN
and increasing the number of additional bonds $B$ leads to localized
states at the band edges and to fairly delocalized states well inside the
band. The increase of $\langle\overline{\chi}\rangle_R$ shown in
Fig.~\ref{prob_kk_zoom} is thus mainly due to the localized band edge
states.

\bw
~
\ew

The participation ratio also dominates the transition probabilities
$\langle\pi_{kj}(t)\rangle_R$, which were presented in Fig.~\ref{pikj} in
Sec.~\ref{sec_transprob}. In general, the
$\pi_{kj}(t) = |\langle k | \exp(-i{\bf H}t) | j \rangle|^2$ averaged over
the distinct realizations read:
\be
\langle\pi_{kj}(t)\rangle_R = \frac{1}{R} \sum_r \left| \sum_n e^{-iE_{n,r}t} \
\langle k | \Phi_{n,r} \rangle \langle \Phi_{n,r} | j \rangle \right|^2.
\ee
Under the assumption that the eigenvalues of SWN are non-degenerate, we
obtain for the initial node $j$
\bea
&& \langle\pi_{jj}(t)\rangle_R = \frac{1}{R} \sum_r \Bigg[ \sum_n
\left| \langle j | \Phi_{n,r} \rangle \right|^4 
\nonumber \\
&& + 
\sum_{n\neq n',n'} e^{-i(E_{n,r} - E_{n',r})t} \left| \langle j
| \Phi_{n,r} \rangle \right|^2 \left| \langle j | \Phi_{n',r} \rangle
\right|^2 \Bigg].
\label{transprob}
\eea
The fluctuations for larger $t$ [$t$-dependent sum in
Eq.~(\ref{transprob})] become suppressed due to the ensemble average. As
can be inferred from Figs.~\ref{pikj}(a)-(c), when increasing $B$ from
$B=0$ only
slightly up to $B/N=0.05$, the fluctuations are already strongly
suppressed. Larger values of $B$, see Fig.~\ref{pikj}(d) for $B/N=1$,
result in a very strong peak at the initial node $j$. Hence, the
fluctuations at the other nodes $k\neq j$ become more and more suppressed
in the ensemble average when increasing $B$.

Now, averaging the time-independent term of Eq.~(\ref{transprob}) over all
nodes $j$ one recovers the LTA of $\langle\overline{\pi}(t)\rangle_R$, see
Eqs.~(\ref{pikk_avg}) and (\ref{chi_avg}): 
\be
\frac{1}{N} \sum_{j} \frac{1}{R} \sum_{r} \sum_n
\left| \langle j | \Phi_{n,r} \rangle \right|^4 =
\langle\overline{\chi}\rangle_R.
\ee
In the ensemble average, all nodes $j$ can be considered roughly equal,
thus every node $j$ gives approximately the same contribution to the sum
over $j$ and we get therefore $\langle\overline{\chi}\rangle_R \approx
\frac{1}{R} \sum_{rn} \left| \langle j | \Phi_{n,r} \rangle \right|^4
\approx \langle\pi_{jj}(t)\rangle_R$. Figure~\ref{pikk_lim} shows that for
increasing $B$ the LTA $\langle\overline{\chi}\rangle_R$ is always larger
than $(2N-2)/N^2$ (the corresponding value for the ring), also leading to
the almost regular shape of the transition probabilities
$\langle\pi_{kj}(t)\rangle_R$ shown in Fig.~\ref{pikj}. 
As noted earlier, single realizations may still show strong interference
patterns. For QG, Kottos and Schanz have given conditions for finding
almost scarred eigenfunctions (states with excess density near unstable
periodic orbits of the corresponding classical chaotic system)
\cite{kottos2003}.  In combination with Smilanskys work on discrete QG
\cite{smilansky2007}, it might be possible in the future to obtain similar
conditions for the networks considered here. 

We stress again that there is no Anderson localization in our system.
Although the states are localized for large $B$, there is still a
non-vanishing transition probability to go from the initial node $j$ to
all other nodes.  Thus, the additional bonds in the SWN do not prohibit
the transport through the network completely, but just hinder it.  Adding
disorder to our system will essentially result in the model considered in
Ref.~\cite{giraud2005}. In this work, the Anderson model was augmented by
additional bonds, such that a SWN develops, which lead to the
localization-delocalization transition.

\section{Conclusion}

We modeled the quantum mechanical transport of (coherent) excitons on
small-world networks by continuous-time quantum walks and computed the
ensemble average of the transition probability to go from one node of the
network to any other node. The transport through the network turns out to
get faster with increasing the number of additional bonds. Distinct from
the classical case, however, where the information of the initial node is
quickly lost, quantum mechanically this information is preserved. During
its time development the exciton is on average most likely to be found at
the initial node. The reason for this is to be found in the network's
eigenstates, which are localized at the band edges, whereas they are quite
delocalized inside the band.

\section*{Acknowledgments.}

Support from the Deutsche
For\-schungs\-ge\-mein\-schaft (DFG), the Fonds der Chemischen Industrie and
the Ministry of Science, Research and the Arts of Baden-W\"urttemberg (AZ:
24-7532.23-11-11/1) is gratefully acknowledged.

\end{document}